\documentclass[10pt,conference]{IEEEtran}

\usepackage[T1]{fontenc}
\usepackage[utf8]{inputenc}
\usepackage{cite}
\usepackage{amsmath}
\usepackage{booktabs}
\usepackage{array}
\usepackage{graphicx}
\usepackage{listings}
\usepackage[dvipsnames]{xcolor}
\usepackage{url}
\usepackage[hidelinks]{hyperref}   %
\hypersetup{pdftitle={CratesBump: A Version-Bump Dataset for the Rust Ecosystem},pdfauthor={Muhammad Hassnain}}
\usepackage[capitalise,nameinlink]{cleveref}  %
\crefname{section}{Section}{Sections}
\crefname{subsection}{Section}{Sections}
\crefname{table}{Table}{Tables}
\crefname{figure}{Fig.}{Figs.}

\newcommand{\dataname}{CratesBump}         %
\newcommand{\toolname}{\textsc{semver-extract}}    %
\newcommand{\nCrates}{219{,}955}                    %
\newcommand{\nPairs}{2{,}079{,}790}                    %

\newcommand{\todo}[1]{}

\newcommand{\hassnain}[1]{}

\begin{document}

\title{\dataname: A Version-Bump Dataset for the Rust Ecosystem}

\author{\IEEEauthorblockN{Muhammad Hassnain}
\IEEEauthorblockA{University of California, Davis\\
Davis, CA, USA\\
mhassnain@ucdavis.edu}}

\maketitle

\begin{abstract}
We release \dataname{}, the first public dataset for
predicting version bumps in the Rust ecosystem, together with
\toolname{}, the tool that builds it. \dataname{} is built from every crate
on crates.io with at least two releases before 14 September 2026 (about
220{,}000 crates), giving over two million consecutive release pairs.
Each pair has 41 features (code changes from AST diffs, dependency
changes, complexity, release cadence, GitHub commit and issue activity),
its source repository where listed, and two labels: the literal reading of
\texttt{MAJOR.MINOR.PATCH}, and Cargo's rule, under which a \texttt{0.y}
bump is a breaking change. \toolname{} regenerates the dataset from public crates.io 
data and the GitHub API. To illustrate its use we (i)~benchmark bump predictors on held-out repositories,
finding the task learnable but far from solved, with lower scores on
later releases; (ii)~measure which feature groups carry
the signal; (iii)~check, against declared dependency ranges, how often
Cargo's rule shields dependents from pre-1.0 breaking releases; and
(iv)~audit a sample of releases with \texttt{cargo-semver-checks}, finding
minor and patch releases that break the public API despite their version
number. Dataset and tool are released under open licenses.
\end{abstract}


\section{Introduction}
\label{sec:intro}
\begin{figure*}[t]
\centering
\includegraphics[width=\textwidth]{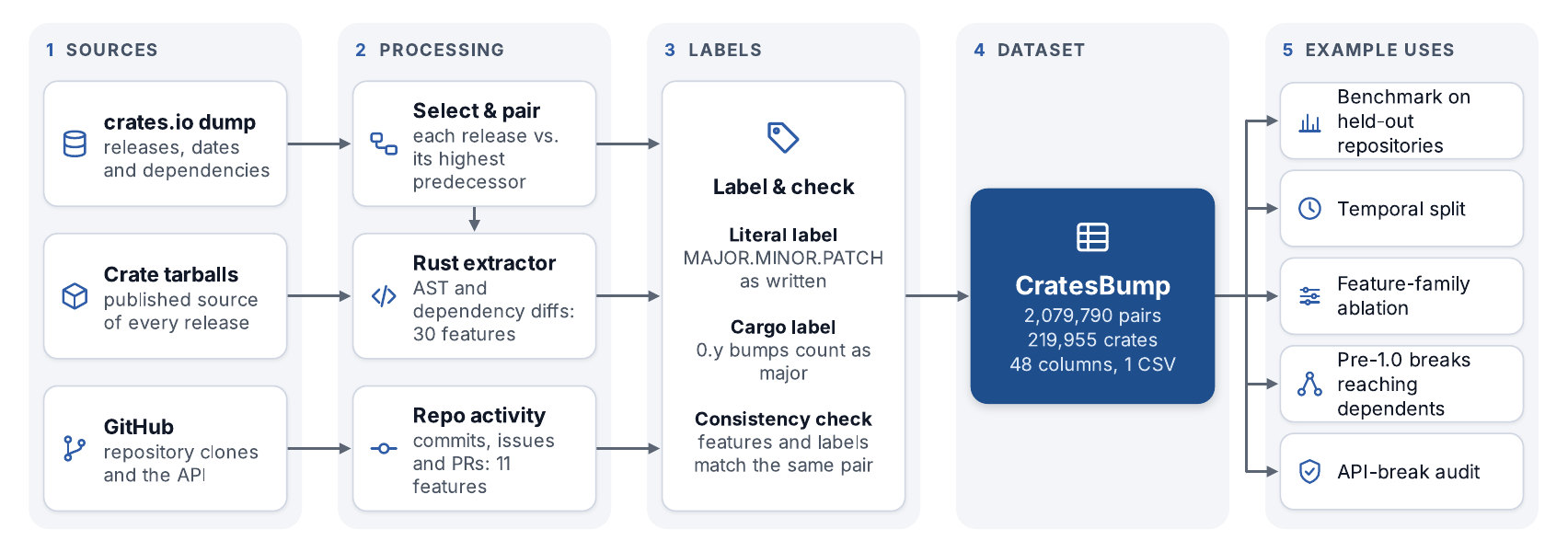}
\caption{How \dataname{} is built (1--4) and used in
\cref{sec:usage} (5).}
\label{fig:workflow}
\end{figure*}
Semantic Versioning (SemVer)~\cite{semver2} is a versioning scheme in
which every release is numbered \texttt{MAJOR.MINOR.PATCH}: a major
release makes breaking changes, a minor release adds backward-compatible
features, and a patch release fixes bugs. Package
managers rely on these numbers to decide whether an upgrade is safe.
The specification makes no promise for \texttt{0.y.z}
versions. Cargo, Rust's package manager, fills that gap~\cite{cargobook}:
the \emph{leftmost non-zero number} marks the breaking boundary, so
\texttt{0.4.5 $\rightarrow$ 0.5.0} is a major bump and
\texttt{0.4.5 $\rightarrow$ 0.4.6} a minor one (npm's caret ranges do the
same~\cite{nodesemver}).

Predicting the bump from the changes between two releases has been
studied as a supervised learning problem on npm~\cite{abdalkareem2022}
and Ansible Galaxy~\cite{opdebeeck2021}, but to our knowledge no public
bump-prediction dataset exists for Rust. Rust's SemVer
checkers~\cite{cargosemverchecks,rustsemverver} answer a related question:
whether the public-API changes they detect are breaking, and so whether a
release's bump is large enough. Both releases must compile, and changes in
behaviour are invisible to them.

This paper contributes:
\begin{itemize}
  \item \textbf{\dataname{}} (\cref{sec:dataset}): \nPairs{}
        release pairs from \nCrates{} crates with at least two releases,
        each with 41 features, its source repository, and both labels;
        and \textbf{\toolname{}} (\cref{sec:tool}), the
        open-source tool that rebuilds it.
  \item \textbf{Reference usage} (\cref{sec:usage}): we benchmark
        bump predictors on held-out repositories, where tree ensembles
        reach a macro-F1 of 0.45--0.46 (baseline 0.24)
        and drop to 0.40--0.42 on later releases. We measure which feature
        families carry the signal and show how pre-1.0 breaking releases
        reach dependents. Finally, we audit releases with
        \texttt{cargo-semver-checks}, checking every flag by hand: at
        least 14.9\% of checked minor and 8.7\% of checked patch releases
        break the public API.
\end{itemize}

\section{Related Work}
\label{sec:related}

\textbf{Datasets of versions and breaking changes.} Several public
datasets record how packages change from one version to the next, and
most of them cover Java. Raemaekers et al.~\cite{raemaekers2013maven}
collect Maven library versions with code metrics. Benelallam et
al.~\cite{benelallam2019maven} model Maven Central's dependencies as a
temporal graph. Jezek and Dietrich~\cite{jezek2017api} build a synthetic
benchmark of API changes, and Reyes et al.~\cite{reyes2024bump} collect dependency updates
that break the build.
Beyond Java, Pinckney et al.~\cite{pinckney2023npmfollower} track every
npm package version, and Rahman et al.~\cite{rahman2026chrono} record resolved
dependencies at releases of selected npm, PyPI and crates.io packages. For Rust, Schueller
et al.~\cite{schueller2022rust} release developer, dependency and
download data. None of these datasets pairs consecutive releases with change
features and bump labels.

\textbf{SemVer practice.} Raemaekers et al.~\cite{raemaekers2017} found
that about a third of Maven releases introduce a breaking change, while
Ochoa et al.~\cite{ochoa2022} found $83.4\%$ of upgrades compliant; in Go,
$28.6\%$ of non-major upgrades break the API~\cite{li2023go}. Bogart et
al.~\cite{bogart2021breaking} compare breaking-change policies across 18
ecosystems and Dietrich et al.~\cite{dietrich2019versioning}
dependency-versioning practices across 17, both including Cargo, and Decan
and
Mens~\cite{decan2021,decan2020magic,decan2021zero} find that package
managers are more permissive about \texttt{0.y.z} than the specification.

\textbf{Bump prediction.} Abdalkareem et al.~\cite{abdalkareem2022}
predict the release type of npm releases from 41 features (change types,
complexity, development activity) over about $6{,}260$ releases of 31
packages. Opdebeeck et
al.~\cite{opdebeeck2021} frame the same task for $81$k Ansible role version
increments, and Lam et al.~\cite{lam2020semantics} call for ``semantic
version calculators'' that tell whether an upgrade is safe. \dataname{}
brings this task to a whole ecosystem.

\textbf{SemVer checkers.} \texttt{cargo-semver-checks}~\cite{cargosemverchecks}
and the now deprecated \texttt{rust-semverver}~\cite{rustsemverver}
compare the public APIs of two builds, as APIDiff~\cite{brito2018apidiff}
does for Java. They detect breaking API changes, but only the kinds their
rules cover; in Java, Sembid~\cite{zhang2022sembid} finds 2--4
times as many APIs with behaviour breaks as with signature breaks. Nowak and
Gruevski~\cite{nowak2023semver} ran \texttt{cargo-semver-checks} on popular
crates.

\textbf{Rust ecosystem.} Kikas et al.~\cite{kikas2017structure} and Decan
et al.~\cite{decan2019} compare Cargo's dependency network with those of
other ecosystems. Studies of crates.io show that unstable compiler
features can break dependent crates~\cite{li2024ruf} and catalogue why
releases are yanked, accidental breaking changes among
them~\cite{yankedrust}. Even code written entirely in safe Rust can harm
its users~\cite{hassnain2024counterexamples}, and Cargo
Sherlock~\cite{hassnain2026sherlock} scores supply-chain trust from
crates.io metadata and analysis-tool results. None of these works labels
version bumps; to our knowledge, \dataname{} is the first public
bump-prediction dataset for Rust.

\section{The \dataname{} Dataset}
\label{sec:dataset}
\subsection{Source and scope}
\label{sec:source}
We use the official crates.io database dump of 12 October 2026
(\cref{fig:workflow}), which
lists every crate and release ever published, for crate selection, release
lists, timestamps and the dependency graph. The dataset is a snapshot at
14 September 2026: it contains every crate with at least two non-yanked
releases published before that instant: 225{,}799 of the dump's
349{,}979 crates (64.5\%), which account for \textbf{99.4\% of all
crates.io downloads}. We also keep crates that never reached 1.0,
since Cargo's rule gives their \texttt{0.y.z} bumps a clear meaning.
They make up 84\% of these crates and carry 62\% of all
downloads. \nCrates{} crates contribute at least one release pair; the
others form no pair (e.g.\ only backports), have no tarballs left (139),
or lost all their pairs to the family cap.

\subsection{Release pairs and features}
\label{sec:extraction}
For each crate, we go through its non-yanked releases in publish order.
Each release numbered above all earlier ones is paired with the highest
earlier release; a \emph{backport} (numbered below an existing release)
is skipped. Publishing \texttt{1.0.0}, \texttt{1.1.0}, \texttt{1.0.5},
\texttt{1.2.0} thus yields \texttt{1.0.0$\rightarrow$1.1.0} and
\texttt{1.1.0$\rightarrow$1.2.0}, so each diff covers exactly the changes
its version number describes. A
final release after its own pre-release (\texttt{1.0.0} after
\texttt{1.0.0-rc.1}) forms no pair, since it has no bump class, but becomes
the new highest version, so \texttt{1.0.1} is compared with \texttt{1.0.0}.
2.6\% of pairs have a pre-release endpoint. So that crates with
thousands of releases and generated crate families (\texttt{aws-sdk-*},
\texttt{solana-*}) do not dominate, we keep at most 100 pairs per crate
and 5{,}000 per name family (the name up to its first \texttt{-} or
\texttt{\_}), sampled with a fixed seed.

For every pair we download both \texttt{.crate} tarballs and compute 41
features (\cref{tab:features}), following Abdalkareem et
al.'s design for npm~\cite{abdalkareem2022}: change-type counts from
\texttt{syn} syntax-tree diffs of every \texttt{.rs} file,
\texttt{Cargo.toml} dependency changes, complexity and size deltas, days
since the previous release, and GitHub activity between the two
releases. Commit features come from a clone of the crate's repository,
issue and pull-request features from the GitHub API, so they reflect the
repository as it was when crawled (October--November 2026). 84.0\% of
pairs have at least one non-zero GitHub feature. The others are all zero
because the crate lists no repository (5.2\%) or one outside GitHub
(3.9\%), its GitHub repository was deleted or made private (4.0\%), or
nothing in the repository falls in the window (2.9\%: half quick
re-releases, half histories that miss the release).\footnote{Two known bugs, to be fixed in the next release, add
to the 4.0\%: unusual GitHub URLs (14 pairs) and unchecked
\texttt{git log} failures (${\le}49$ repos).}

\begin{table}[t]
\centering
\footnotesize
\setlength{\tabcolsep}{3pt}
\caption{The 48 columns of \dataname{}, one row per release pair
(\texttt{prev\_version} $\rightarrow$ \texttt{version}). The 41 feature
columns (lower part) describe what changed between the two releases.}
\label{tab:features}
\begin{tabular}{@{}l r l >{\raggedright\arraybackslash}p{0.50\columnwidth}@{}}
\toprule
\textbf{Group} & \textbf{Cols} & \textbf{Type} & \textbf{What the columns hold} \\
\midrule
Pair key          & 3  & text   & \texttt{package}, \texttt{prev\_version}, \texttt{version} \\
Repository        & 1  & text   & \texttt{repository}, e.g.\ \texttt{github.com/serde-rs/serde} \\
Labels            & 2  & label  & major, minor or patch, by Cargo's rule and by the literal reading \\
Publish time      & 1  & time   & when \texttt{version} was published (UTC) \\
\midrule
Code (AST)        & 20 & count  & files, functions, blocks added/deleted \\
Dependencies      & 4  & count  & dependencies added/removed/updated \\
Complexity, size  & 5  & number & complexity change, lines added/deleted \\
Release timing    & 1  & number & days since \texttt{prev\_version} \\
GitHub commits$^*$ & 7 & number & commits, authors, bug-fix commits \\
GitHub issues$^*$ & 4  & count  & issues and PRs opened/closed \\
\bottomrule
\multicolumn{4}{@{}l@{}}{\rule{0pt}{2.2ex}$^*$Counted between the two publish times; 0 when no GitHub data exists.}
\end{tabular}
\end{table}

\subsection{Bump labels: literal reading vs.\ Cargo's rule}
\label{sec:cargo-relabel}
The literal label \texttt{release\_type\_naive} compares the three numbers
as written; the Cargo label \texttt{release\_type} applies Cargo's rule,
under which the leftmost non-zero component marks a breaking change. Both
labels compare \texttt{version} with \texttt{prev\_version}. We apply a
consistency check against the crates.io dump, which confirms that, for
every pair, the features and labels describe the same two releases. The
labels coincide when MAJOR $\ge 1$ and differ on every
pair whose new version is below 1.0.0 (82.3\% of pairs).
\cref{tab:mechanism} breaks the differences down. Cargo labels are
35.4 / 55.2 / 9.3\% major / minor / patch, against
2.8 / 32.9 / 64.3\% for literal labels.

\begin{table}[t]
\centering
\footnotesize
\caption{Pairs on which the literal and Cargo labels disagree, by kind of
version change: all pre-1.0, 1{,}712{,}585 pairs (82.3\%).}
\label{tab:mechanism}
\begin{tabular}{@{}lrr@{}}
\toprule
\textbf{Version change: literal $\rightarrow$ Cargo label} & \textbf{Pairs} & \textbf{Share} \\
\midrule
\texttt{0.y.z}$\rightarrow$\texttt{0.y.(z+1)}: patch $\rightarrow$ minor  & 1{,}033{,}349 & 60.3\,\% \\
\texttt{0.y.z}$\rightarrow$\texttt{0.(y+1).0}: minor $\rightarrow$ major  & 555{,}772 & 32.5\,\% \\
\texttt{0.0.z}$\rightarrow$\texttt{0.0.(z+1)}: patch $\rightarrow$ major  & 109{,}239 & 6.4\,\% \\
\texttt{0.0.z}$\rightarrow$\texttt{0.y.z}: minor $\rightarrow$ major      & 14{,}225 & 0.8\,\% \\
\bottomrule
\end{tabular}
\end{table}

\noindent\textbf{Storage and schema.} \dataname{} is one gzip-compressed
UTF-8 CSV file (\texttt{data/cratesbump.csv.gz}) with one row per release
pair. Rows are identified by \texttt{package} and \texttt{version}, written
as in the crates.io dump, so users can link them to crates.io's own data
such as downloads or dependencies. Only \texttt{repository} can be empty.
The crate-to-repository map, the audit results, the build reports and
\texttt{SCHEMA.md}, which describes every column, ship alongside.

\section{The \toolname{} Extraction Tool}
\label{sec:tool}
\toolname{} is the open-source pipeline in \cref{fig:workflow}; it
lives in the \texttt{CratesBump} repository~\cite{repo}. Its Rust part
diffs the \texttt{syn} syntax trees and \texttt{Cargo.toml} dependencies
of both tarballs of a pair; Python scripts pair the releases, attach
labels and repositories, and add the GitHub features. A \texttt{Makefile}
runs everything (\cref{sec:availability}). It needs stable Rust,
Python 3.11+, the crates.io dump and a GitHub token.

\toolname{} is built for whole-ecosystem runs on one machine. It reads
release lists from the dump rather than the rate-limited API, so it only
downloads tarballs; 16 parallel workers write one CSV per crate, and a
restart skips finished crates. The GitHub step queries 25 repositories
per GraphQL call and caches every response. A full rebuild takes about
two days, mostly GitHub crawling under its hourly quota; a later snapshot
needs only a newer dump and cutoff date.

\section{Reference Usage}
\label{sec:usage}
\subsection{Benchmark on held-out repositories}
\label{sec:bench}
We train seven classifiers to predict the Cargo label from the
41 features and compare them with two baselines that ignore the
features: one always predicts ``minor'', the most common label; the other
picks each label at random, as often as it occurs in the data.

To test on unseen projects, we hold out 20\% of the source
\emph{repositories}, so crates from one repository are all in training or
all in test (crates without one form their own group). We repeat this
with five random splits and balance only the training data (SMOTE).

We tuned each model by random search (30 settings, 3-fold
cross-validation by repository) on about 500{,}000 pairs. This sample
overlaps the test repositories, but the best tuned model scores no higher
than the best untuned one (both 0.459). We report macro-F1 and the
Matthews correlation coefficient (MCC; 0 for random guessing, even with
imbalanced classes).

\begin{table}[t]
\centering
\footnotesize
\caption{Predicting the Cargo bump label (major, minor or patch) of
pairs from repositories not seen in training: mean $\pm$ s.d.\ over five
repository splits. Macro-F1 averages the three labels' F1; MCC is 0 for
random guessing and 1 for perfect prediction. The first two rows ignore
the features; the best score is in bold.}
\label{tab:baseline}
\begin{tabular}{@{}lcc@{}}
\toprule
\textbf{Model} & \textbf{Macro-F1} & \textbf{MCC} \\
\midrule
Always ``minor'' (baseline)     & 0.236 $\pm$ 0.007 & 0.000 $\pm$ 0.000 \\
Random by label share (baseline) & 0.333 $\pm$ 0.001 & 0.000 $\pm$ 0.001 \\
\midrule
Random Forest             & 0.448 $\pm$ 0.007 & 0.216 $\pm$ 0.012 \\
LightGBM                   & 0.447 $\pm$ 0.007 & 0.223 $\pm$ 0.012 \\
XGBoost                    & 0.457 $\pm$ 0.004 & 0.239 $\pm$ 0.011 \\
Hist Gradient Boosting     & \textbf{0.459} $\pm$ 0.005 & \textbf{0.241} $\pm$ 0.011 \\
Gradient Boosting          & 0.449 $\pm$ 0.005 & 0.225 $\pm$ 0.014 \\
Decision Tree              & 0.437 $\pm$ 0.008 & 0.192 $\pm$ 0.011 \\
Logistic Regression        & 0.361 $\pm$ 0.036 & 0.144 $\pm$ 0.023 \\
\bottomrule
\end{tabular}
\end{table}

The five tree ensembles reach macro-F1 0.447--0.459 (MCC 0.22--0.24),
well above both baselines (\cref{tab:baseline}); patch releases, 9\%
of pairs, are the hardest class (F1 about 0.17). We also find that holding out single crates instead of
whole repositories inflates scores, because crates of one repository
share its GitHub features and release dates. The more a model can
memorise, the larger the effect: the tuned Random Forest rises from 0.448
to 0.461, but with the original, deeper settings from 0.455 to 0.612, so
splits should group by \texttt{repository}.

\subsection{Temporal split}
\label{sec:temporal}
Trained on repositories whose pairs all predate 22 June 2026 (the 80th
percentile of release times) and tested on those whose pairs all follow
it (927{,}421 and 167{,}154 pairs), the tree models reach macro-F1
0.399--0.423, below their 0.437--0.459 on held-out repositories. The test
repositories are young, so the drop mixes a shift over time with one
towards less mature crates.

\subsection{Which feature families carry the signal?}
\label{sec:ablation}
On the first held-out split (Random Forest, macro-F1 0.451), we shuffle
each feature family in the test set five times and record the mean
macro-F1 drop: code change types $+0.029$,
dependencies $+0.014$, complexity and size $+0.012$, days since the
previous release $+0.011$, GitHub commits $+0.008$, and GitHub issues and
pull requests $+0.004$. As on npm, where change types were the strongest
predictors~\cite{abdalkareem2022}, code features carry the most signal and
GitHub features the least, although the
original unbounded-depth Random Forest leaned on GitHub more ($+0.046$,
$+0.027$).

\subsection{What lets pre-1.0 breaking releases reach dependents?}
\label{sec:downstream}
A release such as \texttt{0.4.5 $\rightarrow$ 0.5.0} looks minor but is
breaking under Cargo's rule. Do the crates that depend on it pick it up
automatically? Usually not: by default, a dependency on \texttt{0.4.5}
accepts updates only below \texttt{0.5.0}, and this default covers
94.9\% of the 33.6\,M regular dependency declarations in the crates.io
dump. For each of the 569{,}997 such releases in \dataname{}, we check
whether the requirement of every crate that depended on the package just
before the release accepts it (ignoring pre-release tags). Only
26{,}558 of 1{,}200{,}628 dependency links (2.2\%) would have picked up
the breaking release, and only 8{,}525 of these releases (1.5\%) reached
any dependent. Almost all such links use loose requirements: any
\texttt{0.y.z} version (e.g.\ \texttt{0.*}; 58\%), no upper bound (e.g.\
\texttt{>=0.22}; 24\%) or any version (\texttt{*}; 15\%).

\subsection{External API audit}
\label{sec:audit}
Do minor and patch releases break their public API? We checked a random
sample with \texttt{cargo-semver-checks}~\cite{cargosemverchecks} 0.51.0,
which compares the public APIs of two versions using rules called lints
and says which kind of release the changes require. We drew 450 pairs,
150 per Cargo label and each from a different crate, and built them
offline in a container.

The tool could check 305 of them. The rest are command-line programs or
macro crates it cannot read (71), or no longer build or resolve (74; eight only
because the tool turns on conflicting features). Older releases fail more
often.

The tool flagged 21 of 101 minor and 10 of 103 patch releases as needing
a major release. We read the code of all 31: 24 are real breaks, 6 are
unlikely to matter (e.g.\ generated bindings), and 1 is not a break. So
\textbf{at least 14.9\% of minor and 8.7\% of patch releases break their
public API} (95\%~CIs 9.2--23.1\% and 4.7--15.8\%); in 20 of the 24,
ordinary code using the crate stops compiling.

Some caveats. The rates are lower bounds, since the tool misses behaviour
changes and most type changes. Most checked minor releases (92 of 101)
are \texttt{0.y.z}, so the minor rate mostly describes pre-1.0 crates. One
break is a pre-release; without pre-releases the rates are 14.3\% and
9.0\%. With one release per crate, the rates describe a typical crate,
not all releases.

Our rates are well above the 3.2\% that Nowak and
Gruevski~\cite{nowak2023semver} found with an early version of the tool
on the 1{,}000 most-downloaded crates (ours combined: 11.8\%). The newer
lints explain little of the gap: with only the 45 lints of mid-2023, our
rates drop to 12.9\% and 6.8\%. Their study also used default features
only, confirmed each break with failing code, and checked every release
of popular crates; we have not tested which difference matters.

\section{Limitations and Future Work}
\label{sec:limits}
\textbf{Labels and GitHub features.} Labels are declared bumps: Cargo's
rule fixes how a version change is read, not whether the maintainer chose
correctly (\cref{sec:audit}). GitHub features reflect each repository as crawled and are shared
by its crates, hence the held-out-repository splits. A zero can mean ``no data'' or
``no activity'' (\cref{sec:extraction}).

\textbf{Caps, snapshot and analyses.} Capped crates contribute a random
sample of their pairs (\toolname{} can re-extract them in full), and
releases from 14 September 2026 on are missing. The audit covers only
library crates that still build, and the dependents analysis builds
nothing.

\textbf{Future research.} \dataname{} can help study which changes
maintainers under-version and whether a model can flag them before
publishing, when and why crates leave \texttt{0.y.z}, and whether bump
predictors transfer across ecosystems and time.

\textbf{Future improvements.} Later releases could add a
\texttt{cargo-semver-checks} label per buildable pair, regular snapshots,
changelog and test signals, non-GitHub hosts and a missing-data flag.

\section{Availability}
\label{sec:availability}
\dataname{} and \toolname{} are on GitHub~\cite{repo} as release
\texttt{v1.0.0}; the code is under the MIT license and the dataset under
CC~BY~4.0. The repository's \texttt{Makefile} regenerates every table and
number in this paper except three GitHub-coverage details that need the
crawl caches. The benchmark and temporal split need only the CSV file; the
other analyses also need the crates.io dump, a GitHub token and, for the
audit, Docker. Rebuilding the dataset from scratch
(\texttt{make rebuild-full}) takes about two days.

\section*{Acknowledgment}
The author used Claude Code (Anthropic) to help write the code and edit
the text, and checked all of it.

\bibliographystyle{IEEEtran}
\bibliography{ref}

\end{document}